# Asymmetrical Path Interference Test of Light
## -----An adjudicative test about light's essence

## Mei   Xiaochun

(Department of Physics, Fuzhou University, E-mail: mxc001@163.com Tel:0086-591-87614214)

(N0.27-B, South Building, Zhongfu West Lake Gardern, Fubin Road, Fuzhou, 350025, China)

**Abstract** The light's asymmetrical path interference test is put forward in the paper. In the test, two different results would arise under the same experimental conditions if light is regarded as wave or particle. Therefore, the test can help us to comprehend which concept, wave or particle, is more essential for micro-particles. Taking advantage of the test, we can prove that it is impossible for a single photon to pass through two slits simultaneously to achieve self-interference. Perhaps the test would expose a very important result that the overlap of coherency waves is not a necessary or essential condition to produce the light's interference as shown in the so-called "the ghost interference of light" [1] .

The wave-particle duality is the most perplexed paradox in modern physics. Because it can't be explained rationally up to now, people seems becoming used to it. Now physicists have accepted this duality and think that a micro-particle can be regarded as both wave and particle simultaneously. However, wave always spreads over the whole space but particles always occupy small volumes. Those two pictures are completely contradictory in essence. In order to explain the duality, Bohr putted forward the principle of complementary, thought that the pictures of particle and wave are complementary for micro-particles. The experiments also show that under same experimental conditions particle's natures appear, but under other experimental conditions wave's natures appear for micro-particles. Being wave or particle depends on the selections of experimental conditions. Because these two kinds of experimental conditions can't exist simultaneously, the duality contradiction seems to be avoided in surface. But the problem is not so simple. If the concepts both wave and particle are completely equivalent for micro-particles in logic, the problems existing in the current explanation of quantum mechanics would not be eliminated forever.

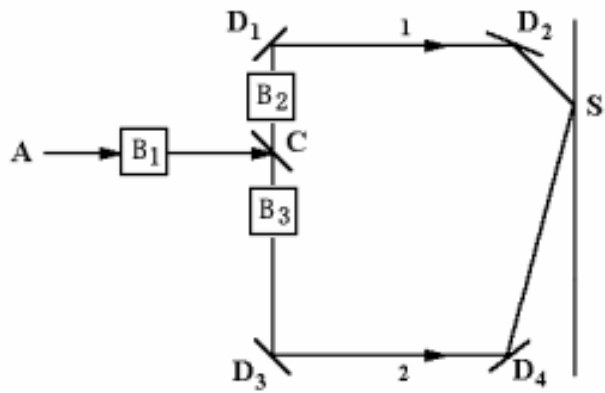

Fig.1 Light's asymmetrical path interference test

The Light's asymmetrical path interference test is put forward in the paper. The key of the test is that when light is regarded as wave or particle, under the same experimental condition, two different results would arise in theory. In this case, light's wave-particle duality can be eliminated through the experiments. Therefore, the test can help us to judge which concept wave or particle is more essential for micro-particles. Meanwhile, By means of this test, we can prove that it is impossible for a single photon to pass through two slits simultaneously to achieve self-interference.

As shown in the figure above, $A$ is the optical source. $B_1$, $B_2$ and $B_3$ are the same light switches, for example the Pockels' cell which can operate quickly to close or open the light's passageway. C is the beam splitter. The distances from $C$ to $B_2$ and $B_3$ are the same and arbitrary, for example taking $0.1m$. The distance along Path 1 from $B_2$ via the mirrors $D_1$ and $D_2$ to the screen $S$ is 4m and along Path 2 from $B_2$ via $D_3$ and $D_4$ to the screen is 7m. The difference of optical paths is $L_1 = 3m$. At beginning, all three optical switches are turned on so that the light can pass continuously. If proper laser is chosen, for example the He-Ne laser with coherency length $L_2 > 3m$, the interference fringes will be formed on the screen $S$ by the superposition of light beams coming from Path 1 and 2.

Then we turn off and turn on the switch $B_1$ successively but keep $B_2$ and $B_3$ in open states. In this way, continuous light is transformed into pulses. Suppose that the open time interval of light switches is chosen to be $\Delta t_1 = 9 \times 10^{-9} s$, during this period of time the traveling distance of light is $L_3 = c\Delta t_1 = 2.7m$. The closing time interval is chosen to be $\Delta t_2 = 10^{-8} s$, during this period of time the traveling distance of light is $L_4 = c\Delta t_2 > 3m$. In this way, only after the first pulse moving along Path 2 has reached the screen, the second pulse moving along Path 2 begins to reaches the screen, so that the two former and latter pulses do not overlap on the screen.

Now let us analysis the results in theory. If the light is regarded as a wave, the largest length of the pulse wave train is 2.7m when the switch $B_1$ operates. After a wave train passes through $C$, it splits up into two beams. Then two beams travel along Paths 1 and 2 individually and simultaneously. When the tail of a wave train moving along Path 1 has reached $S$, the front of another wave train moving along Path 2 is still on the way owing to the fact that the distances along Path 1 and 2 are different. It means that these two wave trains can't overlap on the screen, so that the interference fringes can't be formed. Or speaking simply, after the light switch $B_1$ operates, the coherency length of light is changed from $L_2 > 3m$ into $L_3 \leq 2.7m$ (The coherency length of light is usually defined as the length of wave train in classical optics.). Therefore, the coherency condition of light cannot be satisfied and the interference fringes would disappear after the light switch $B_1$ operates. This is just the result of classical optics and there is no surprising to this result from the angle of classical theory to regard light as a continuous wave.

But on the other hand, if light's essence is regarded to be particles or photons, the effect of light switch $B_1$ is only to let photons to pass or obstruct them (actually deflects photons to other directions by means of the Pockels' cell). The formula of coherency length is $L_2 = \lambda^2 / \Delta\lambda = h^2 c / \Delta E$, in which $\lambda$ and $E$ are the wavelength and energy of photons individually. For those photons that have passed through the switch, the operation of switch $B_1$ does not change their natures, i.e., light's wavelength, energy, polarization and so on do not change. Therefore, $\Delta\lambda$ and $\Delta E$ are unchanged, so the coherency nature of photons is still unchanged after the switch $B_1$ operates. In this case, the photons traveling along two different paths are still coherent ones. The result is that the original interference fringes on the screen would be unchanged after switch operates. In fact, a great many of experiments have shown that as long as coherency condition is satisfied, no matter what kind of light, continuous light or pulse light, even photon emitted one by one (i.e., the self-interference of a single photon), the interference pictures are the same.

Therefore, in the test of light's asymmetrical path interference, we can't judge the experiment results in theory if the concepts of particle and wave are equivalent. The practical experiment is needed to determine whether interference picture would disappear or not after light switch operates. If interference picture does not change, the essence of light should be particle and the wave nature of light is just an apparent behavior. If interference picture disappears, the essence of light should wave and the particle's nature of light is only an apparent behavior. In this test, the concepts of wave and particle are not equivalent to each other in logic. Because the experimental condition is the same, there exists no wave-particle duality. Only one of them is essential, another is apparent.

We can also use this test to prove that it is impossible for a single photon to pass through double slits simultaneously to achieve self-interference. According to widely accepted idea at present, a single photon can be regarded as a wave. The wave can be split into two sub-waves by beam splitter. Then two sub-waves travel along two paths, interfere each other on the screen and form an integral photon again. This process is just so-called self-interference of a single photon. Suppose that the photons are emitted from the light source one by one. The second photon is emitted after the first one has reached the screen. If a photon corresponds to a wave, a free photon corresponds to a free plane wave with an infinite wave train. In this experiment the photons are not free ones for the existence of interaction. In general, the wave train length is equivalent to the coherency length. So the wave train length of a single photon is also $L_2 > 3m$ at beginning. Then we turn on the switch $B_1$, but operate $B_2$ and $B_3$ synchronously, i.e., turn on and turn off $B_2$ and $B_3$ simultaneously.

After the switches $B_2$ and $B_3$ operate, each wave train of a single photon is cut into two or more short wave with length $L_3 \leq 2.7m$. Similarly, because the lengths of two paths are different, when the tail of a short wave train traveling along Path 1 has reached the screen, the front of another short wave train traveling along Path 2 is still on the way. So these two short wave trains can't overlap on the screen so that an integrated photon can't be formed again. What we can observe on the screen is only two or more fragments of a single photon. However, this is certainly impossible whether or not the interference picture disappears at last. Because what we observe on screen is always an integrated photon, the result notes that it is impossible for a single photon to pass through two paths (or double slits) simultaneously to achieve self-interference. A photon can only travel along one of two paths (or one of double slits) to reach the screen. In the current discussions about a single photon's self-interference, we always suppose that the distances of two paths are nearly equal so that the problem is hidden. By taking advantage of the test, it can be said that the picture of double slit interference of a single photon is not actually caused by the overlap of two wavelets. Or speaking more clearly, a single photon can't split into two waves moving along two paths simultaneously. Therefore, it is necessary for us to look for the new explanation of the double slit interference of a single photon

In fact, the experiments of the so-called "ghost interference" of light (1), achieved by D.V. Strekalov Univer etc. in University of Maryland, has shown that the interference fringes can also arise without the overlap of light waves. In the experiment of "ghost interference", the interference picture is caused by the so-called two photon's entanglement, according to general understanding. The result denotes actually that the overlap of coherency waves is not a necessary or essential condition for the light's interference.

By the discussion above, it can be predicted that after the switch $B_1$ operates (the switches $B_2$ and $B_3$ turn on), the interference picture would still remain, means that the essence of light is particle. In fact, the interference fringes can be explained as the result of statistical average of a large number of photons interacting with environment, as well as the result of entanglement interaction between particles. If the environments are different (for example, the difference between double slit and single slit, two paths and one path), the results are certainly different (diffraction or interference). In these cases, we do not use the picture of wave's overlap, though it is a very simple and effective one. In other words, the result of interaction between particles and environments are equivalent to the overlaps of waves. So we have two equivalent descriptive methods or pictures to explain the light's interference. In the picture of wave's overlap, no interactions are considered. In the picture of interaction, no wave's overlap is considered. But in same special situations, these two descriptions may be unequal to each other just as the experiments shown in this paper and "ghost interference".

So it can be said that for the light's interference, the descriptive method of interaction is more general and essential. It is unnecessary and improper for us to suppose that a photon is also a continuous wave at any instant, thought we can think that the statistical average behavior of a single particle moving in space during a long enough period of time can be equivalent to a wave. In fact, it is well known that the wave of micro-particles is a probability wave, not a classical material wave that continuously distributes over whole space. It should be emphasized that in the description of micro-particle's probability wave, the particle's concept is more essential. In fact, the wave is a macro-concept but the particle is a micro-concept, both can't be equal to each other directly. The situation is the same as that in classical physics. It is meaningless for us to talk about a single macro-particle's temperature and pressure in classical physics. Only for a macro-system with a large number of particles, the concepts of temperature and pressure are meaningful.

The essence of wave particle duality is the most foundational problem in the explanation of quantum mechanics. The light's asymmetrical path interference test would be useful for us to provide a clue to understand microcosm deeply.